\documentclass[aps,prb,showpacs,groupedaddress,twocolumn]{revtex4}
\usepackage{graphicx}
\usepackage{amsmath}
\usepackage{bm}
\usepackage{xcolor}

\begin{document}

\title{Nonlocal transport in a hybrid two-dimensional topological insulator}

\author {Yanxia Xing$^{1,*}$ and Qing-feng Sun$^{2,3,\dagger}$}

\affiliation{
$^1$Department of Physics, Beijing Institute of Technology, Beijing 100081, China \\
$^2$International Center for Quantum Materials, Peking University, Beijing 100871, China\\
$^3$Collaborative Innovation Center of
Quantum Matter, Beijing, China}

\begin{abstract}
We study nonlocal resistance in an H-shaped two-dimensional HgTe/CdTe quantum well consist of injector and detector, both of which can be tuned in the quantum spin Hall or metallic spin Hall regime.
Because of strong spin-orbit interaction, there always exist spin Hall effect and the nonlocal resistance in HgTe/CdTe quantum well. We find that when both detector and injector are in the quantum spin Hall regime, the nonlocal resistance is quantized at $0.25\frac{h}{e^2}$, which is robust against weak disorder scattering and small magnetic field. While beyond this regime, the nonlocal resistance decreases rapidly and will be strongly suppressed by disorder and magnetic field. In the presence of strong magnetic field, the quantum spin Hall regime will be switched into the quantum Hall regime and the nonlocal resistance will disappear. The nonlocal signal and its various manifestation in different hybrid regimes originate from the special band structure of HgTe/CdTe quantum well, and can be considered as the fingerprint of the helical quantum spin Hall edge states in two-dimensional topological insulator.
\end{abstract}

\pacs{72.20.-i, 
73.20.-r,   
72.25.Dc,   
73.63.-b}   

\maketitle

\section{introduction}

A topological insulator is a special quantum
matter\cite{Phys.Rev.Lett.95.226801.Kane.2005}. Due to its
particular band structure that the bulk states have a gap and the
surface states can exist in the bulk band gap, the topological insulator
behaves as an insulator in its interior and
behaves as a metal on the surface. This leads to quantum spin Hall (QSH) effect.
Different from magnetic field induced quantum Hall effect where
the time-reversal symmetry is broken, the QSH effect arises from
strong spin-orbit interaction (SOI) and is protected by the time-reversal
symmetry. In the QSH regime, the electron spins are locked to their
momenta and the boundary states are helical, i.e., there are two
time reversed counter propagating edge states occupied by electrons with opposite
spin. Such a topological state can be signed by $Z_2$
index\cite{Phys.Rev.Lett.95.146802.Kane.2005}. In topological insulator,
the boundary states are protected by the time-reversal symmetry and the
backscattering between boundary states is strongly suppressed. As a result,
the helical edge states are robust against the non-magnetic disorder.\cite{PhysicalReviewLetters103.036803.Jiang.2009}
In the bulk energy band beyond the QSH phase, the topological insulator
behaves as a metal, which is different from the conventional
two-dimensional (2D) metal. In conventional metal, the electron wave functions are
localized as long as there exists any weak
disorder\cite{Phys.Rev.Lett.42.673-676.Abrahams.1979}. However, in
topological insulator, the metal state can exhibit quantum
conductance in the moderate disorder, which is called
topological Anderson insulator phenomena\cite{Phys.Rev.Lett.102.136806.Li.2009,Phys.Rev.B84.035110.Xing.2011,Phys.Rev.B80.165316.Jiang.2009}.

Up to now, the topological
states have been predicted theoretically in several materials, such as the HgTe/CdTe quantum
well\cite{Science318.766-770.Konig.2007,Science314.1757-1761.Bernevig.2006},
the InAs/GaSb quantum well\cite{PhysicalReviewLetters100.236601.Liu.2008,PhysicalReviewLetters107.136603.Knez.2011,PhysicalReviewLetters109.186603.Knez.2012}, the monolayer
graphene with intrinsic SOI\cite{Phys.Rev.Lett.95.146802.Kane.2005,Phys.Rev.Lett.95.226801.Kane.2005,Phys.Rev.B75.041401.Yao.2007}, and the gated bilayer
graphene\cite{Phys.Rev.Lett.100.036804.Martin.2008} that contains
the one-dimensional chiral edge states.
In experiment, the HgTe/CdTe quantum well and the InAs/GaSb
quantum well with inverted band have been successfully discovered as 2D
topological insulators with the QSH
phase\cite{Science314.1757-1761.Bernevig.2006}. Since the discovery
of the topological insulator, many works have been concentrated on
the verification of its helical edge states. Such as, Konig and
co-workers\cite{Science318.766-770.Konig.2007} observed a quantized longitudinal conductance at about $2e^2/h$
that is consistent with the number of edge states predicted theoretically.
Then, the transport along the edge states was confirmed\cite{Science325.294-297.Roth.2009,Science325.278-279.Buttiker.2009}.
However, it remained to be shown that whether the transport due to the helical edge
states is spin polarized. For this purpose, Br$\ddot{u}$ne and
co-workers\cite{NaturePhysics8.486-491.Brune.2012} designed another
experiment, in which the QSH effect and metallic spin Hall (MSH)
effect are combined in a single HgTe/CdTe quantum well device
using split gate technique. Through observation of nonlocal
resistance in a H-shaped device, the spin polarization of edge state
is then determined. In the work, in order to estimate the nonlocal resistance, the semiclassical simulation are performed\cite{NaturePhysics8.486-491.Brune.2012}, but it breaks down when the chemical potential is close to the insulating gap. It is nowhere near enough for the QSH system. On the other hand, the nonlocal transport originates from the Hall effect, including the quantum Hall effect and the spin Hall effect. So, we must carefully examine the
nonlocal effect to illustrate the role of the QSH state. Furthermore, as shown in the above experiments, the nonlocal signal deviated from the standard pattern predicted theoretically
because of the various impurity and dephasing effect. Therefore, the detailed mechanism of the
nonlocal transport in HgTe/CdTe quantum well, especially for the nonlocal transport in the QSH regime, is not very clear.

\begin{figure}
\includegraphics[width=6.0cm,totalheight=7.0cm, clip=]{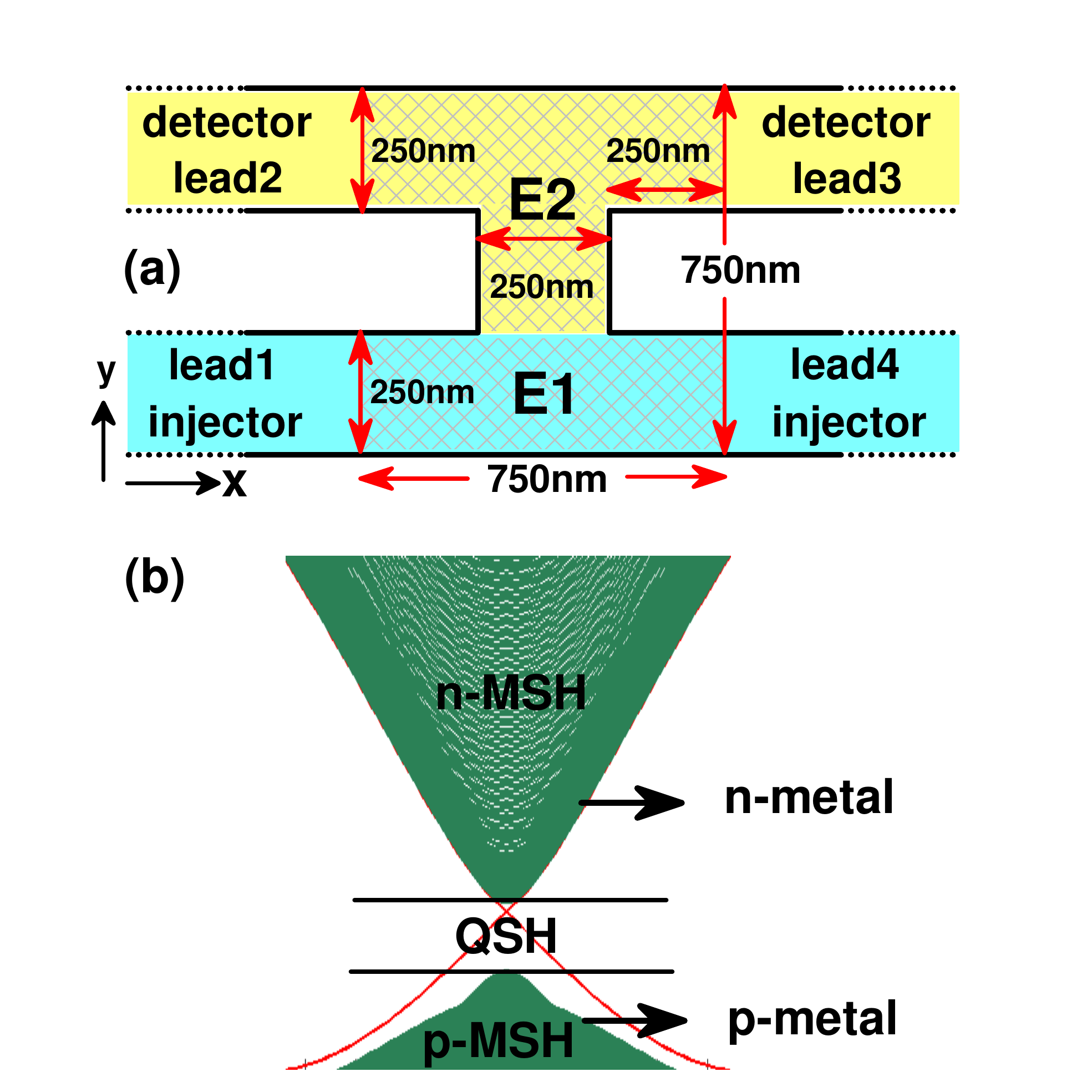} 
\caption{ (Color
online) (a) Sketch of the H-shaped hybrid four-terminal system. Through
spilt gate technique, the on-site energies $E1$ and $E2$ in the injector (green region) and
detector (yellow region) can be tuned separately. In the system,
the scattering region (shadowed region) is connected to four leads. (b) Band structure of HgTe/CdTe:
the conduction band (n-type carriers corresponding to n-MSH effect)
and the valence band (p-type carriers corresponding to p-MSH effect) are
separated by bulk band gap which holds the helical edge states and
consequently the QSH effect.} \label{structure}
\end{figure}

In this paper, based on a four-band tight-binding model and aided by
Keldysh nonequilibrium Green's function, we study the nonlocal
transport in a hybrid HgTe/CdTe quantum well, especially in the QSH regime.
Following the recent experiment by Br$\ddot{u}$ne et.
al.\cite{NaturePhysics8.486-491.Brune.2012}, we consider an H-shaped
device based on the HgTe/CdTe quantum well as shown in Fig.1(a).
The on-site energies $E1$ and $E2$ in the bottom (green) and top
(yellow) regions can be tuned separately by the split gates above the two regions. As a result, the two regions can then be in the QSH or MSH regime [see Fig.1(b)]. When changing $E1$ and $E2$, the quantum well will be in
different regime, i.e., different hybrid structure, including QSH-QSH regime, QSH-MSH regime, and MSH-MSH regime.
When injecting current from the bottom two terminals, spin accumulation or spin polarized potential is
generated in the bridge between the bottom and top terminals due to the spin Hall effect. Then, because of the inverse SHE, the charge voltage will be detected in the top two terminals and leads to the nonlocal resistance in the HgTe/CdTe quantum well. It is obvious that there exists nonlocal resistance in all the hybrid devices because of the spin Hall effect. However, their manifestation is very different.
For example, the nonlocal resistance in QSH-QSH device is quantized with
the value of $\frac{h}{4e^2}$, that is robust against moderate
disorder and weak magnetic field, while the nonlocal resistance in
MSH-QSH or MSH-MSH device is oscillating and fragile. In the presence of
strong magnetic field, the quantized nonlocal resistance will shrink
and disappear finally. All these features of the nonlocal
resistance: the robust quantized nonlocal resistance in the QSH regime,
the fragile nonlocal resistance in the MSH regime, and the vanishing
nonlocal resistance in the presence of strong magnetic field, are the
particular characters of HgTe/CdTe quantum well, which can be
derived from the special band structure of HgTe/CdTe quantum well.
The details will be expatiated in the section III.

The rest of the paper is organized as follows. In Sec. II, based on
the four-band tight-binding representation, the model Hamiltonian of
system including central scattering region and attached leads is
introduced. The formalisms for calculating the Green's functions and
nonlocal resistance are then derived. Sec. III gives numerical
results together with detailed discussions. Finally, a brief summary is
presented in Sec. IV.

\section{model and Hamiltonian}

The whole system we consider is composed of the HgTe/CdTe quantum
well with inverted band. In general, the structure of quantum
well is asymmetry, which leads to the external Rashba SOI. The
Hamiltonian of the system can be written as $H(k) = H_0(k)+H_R(k)$, where
$H_R$ comes from the Rashba SOI, and
\begin{eqnarray}
H_0(k) = \left(
\begin{array}{cc}
H_\uparrow(k) & 0\\
0 & H_\downarrow(k)
\end{array}
\right).\label{Hk}
\end{eqnarray}
From the time-reversal symmetry, we can get $H_{\uparrow}({\bf k})=H^*_{\downarrow}(-{\bf
k})$, where
\begin{eqnarray}
H_{\uparrow}({\bf k}) &=& \hbar v_F({\bf k}_x\sigma_x-{\bf
k}_y\sigma_y)+(m+C{\bf k}\cdot{\bf
k})\sigma_z \nonumber \\
&+&D{\bf k}\cdot{\bf
k}\sigma_0,
\end{eqnarray}
where $\sigma_{x,y,z}$ are Pauli matrices presenting the pseudo spin
formed by $s$ and $p$ orbitals, $\sigma_0$ is the $2\times2$ unitary matrix
in the pseudo spin space. $v_F$ is Fermi velocity. C, D, and m are the system parameters, which can be
experimentally controlled. While the Hamiltonian due to Rashba SOI $H_R$ is expressed as
\begin{eqnarray}
H_R(k) = \left(
\begin{array}{cccc}
0&0&it_Rk_- & 0\\
0&0&0&0\\
-it_Rk_+&0&0&0\\
0&0&0&0
\end{array}
\right)\label{Hkr}
\end{eqnarray}
where $t_R$ is the Rashba SOI strength.

Equations~(\ref{Hk}) and (\ref{Hkr}) are the low-energy effective Hamiltonians of the HgTe/CdTe quantum
well\cite{Science314.1757-1761.Bernevig.2006,Phys.Rev.Lett.101.146802.Liu.2008}
from the $k\cdot p$ perturbation. For an ideal crystal lattice,
i.e., the infinite periodic system, the momentum $k$ is a good
quantum number. In this case, using Hamiltonians (\ref{Hk}) and (\ref{Hkr}) is convenient.
However, here we consider charge transport in the
H-shaped system, Hamiltonian can't be expressed in momentum space,
it should be expressed in real space. To do this, we substitute ${\bf k}$ with $-i\nabla$ and use the
finite-difference approximation, then the total Hamiltonian $H$ is transformed
into the tight-binding Hamiltonian in square lattice. It is given
by\cite{Phys.Rev.Lett.102.136806.Li.2009,Phys.Rev.B80.165316.Jiang.2009}
\begin{eqnarray}
H&=& \sum_{\bf i} d^{\dagger}_{\bf i}
H_{\bf ii} d_{\bf i}\nonumber
\\&&+\sum_{\bf i} d_{\bf i}^{\dagger}
H_{{\bf i},{\bf i}+{\bf a}_x}e^{i\phi_{{\bf i},{\bf i}+{\bf a}_x}} d_{{\bf i}+{\bf a}_x}
+h.c. \nonumber
\\&&+\sum_{\bf i} d_{\bf i}^{\dagger} H_{{\bf i},{\bf i}+{\bf a}_y} d_{{\bf i}+{\bf
a}_y}+h.c.,
\label{Hr}
\end{eqnarray}
where $d_{\bf i}=[d_{s,{\bf
i},\uparrow},d_{p,{\bf i},\uparrow},d_{s,{\bf
i},\downarrow},d_{p,{\bf i},\downarrow}]^T$ with `$T$' denoting transpose, and
$d_{s(p),{\bf i},\uparrow(\downarrow)}$ and $d_{s(p),{\bf i},\uparrow(\downarrow)}^{\dagger}$ are the
annihilation and creation operators for $s$ or $p$ orbital at site ${\bf
i}$ with spin up or spin down, respectively.
${\bf i}=({\bf i}_x, {\bf i}_y)$ is the index of
the discrete site of the system in the square lattice, and ${\bf
a}_x=(a,0)$ and ${\bf a}_y=(0,a)$ are the unit vectors of the square
lattice with $a$ the lattice constant.
In zero magnetic field, the Hamiltonian (\ref{Hr}) possesses
time-reversal invariant. In the presence of a uniform
perpendicular magnetic field ${\mathbf B}=[0,0,B]$, the time-reversal
symmetry is broken. In coulomb gauge, considering the seimi-infinite leads being along the $x$-direction, the
vector potential is set as ${\mathbf A} = [-By,0,0]$ which is $y$ dependent but periodic in the $x$-direction. In this case, a phase $\phi_{\bf ij}$ is
generated in the hopping term $H_{{\bf i},{\bf i}+{\bf a}_x}$.
The phase $\phi_{\bf ij}=\int_{\mathbf i}^{\mathbf j} {\mathbf
A}\cdot d{\mathbf l}/\phi_0$ with flux quanta $\phi_0=\hbar/e$.
In Eq.(\ref{Hr}), $H_{\bf ii}$ and $H_{{\bf i},{\bf i}+{\bf a}_{x}({\bf a}_{y})}$ are all $4\times4$ block matrix
that are expressed as
\begin{eqnarray}
&&H_{\bf ii}= (\epsilon_{\bf i}-\frac{4D}{a^2})(s_0\otimes\sigma_0) +(m-4\frac{C}{a^2})(s_0\otimes\sigma_z)\nonumber \\
&&H_{{\bf i},{\bf i}+{\bf a}_x}=\frac{D}{a^2}(s_0\otimes\sigma_0)+\frac{C}{a^2}(s_0\otimes\sigma_z)\nonumber \\
&&~~~~~~~~~-i\frac{\hbar v_F}{2a}(s_z\otimes\sigma_x)
+ i\frac{t_R}{2a}(s_y\otimes\frac{\sigma_0+\sigma_z}{2})
\nonumber \\
&&H_{{\bf i},{\bf i}+{\bf a}_y}=\frac{D}{a^2}(s_0\otimes\sigma_0)+\frac{C}{a^2}(s_0\otimes\sigma_z)\nonumber \\
&&~~~~~~~~~+i\frac{\hbar v_F}{2a}(s_0\otimes\sigma_y)- i\frac{t_R}{2a}(s_x\otimes\frac{\sigma_0+\sigma_z}{2}),
\end{eqnarray}
where $s_{x,y,z}$ are the pauli matrices denoting the real spin and $s_0$ is the $2\times2$ unitary matrix extented in real spin space. $\epsilon_{\bf i} = E1(E2) +w_{\bf i}$ with $E1$ and $E2$ being the on-site
energies in detector (yellow) and injector (green) regions in
Fig.1(a). $w_{\bf i}$ comes from the disorder effect that is a
random on-site potential which is uniformly distributed in the
region $[-w/2,w/2]$.

Based on above Hamiltonian, the charge current flowing to the $p$-th lead can be
calculated from the zero temperature Landauer-Buttiker
formula\cite{book}
\begin{equation}
J_p = \frac{e^2}{h}\sum_{q}T_{pq}(V_p-V_q)
\end{equation}
where $p,q=1,2,3,4$ are the index of the four leads. $T_{pq}$ is the
transmission coefficient from terminal q to terminal p.
In the following, we derive $T_{pq}$ in the system without and with Rashba SOI.
In the absence of Rashba SOI, the Hamiltonian with spin up and spin down are decoupled, and
the self energy and Green's functions for the system with spin up and spin down can be calculated separately.
Then transmission $T_{pq}=T_{pq,\uparrow}+T_{pq,\downarrow}$ and
$T_{pq,\sigma}(E)=Tr_{r,o}[\Gamma_{p,\sigma}
G^r_{pq,\sigma}\Gamma_{q,\sigma} G^a_{qp,\sigma}]$,
where the "${\rm Tr}_{r,o}$" denotes the trace over real space and orbital space (s and p orbitals),
the line-width function $\Gamma_{q,\sigma} =
i[\Sigma^r_{q,\sigma}-\Sigma^{r,\dagger}_{q,\sigma}]$, and
$G^r_{pq,\sigma}$ is the Green's function matrix whose rows and columns mark
the lattices that are nearest to $p$ and $q$ lead respectively. The
Green's function $G^r_\sigma(E) =
G^{a,\dagger}_\sigma(E)=(EI-H_{\sigma}-\sum_q
\Sigma^{r}_{q,\sigma})^{-1}$, where $H_\sigma$ is Hamiltonian matrix with spin $\sigma$ in
the central region and $I$ is the unit matrix with the same
dimension as that of $H_{\sigma}$, and $\Sigma^{r}_{q,\sigma}$ is the
retarded self-energy function contributed by the electrons with spin $\sigma$ in lead $q$.
The self energy function can be obtained from
$\Sigma^r_{p,\sigma}=H_{cp,\sigma}g^r_{p,\sigma} H_{pc,\sigma}$,
where $H_{cp,\sigma}$ is the coupling from central region to lead
$p$ and $g^r_{p,\sigma}$ is the surface retarded Green's function of
semi-infinite lead $p$ which can be calculated using transfer-matrix
method\cite{Phys.Rev.B23.4997-5004.Lee.1981,Phys.Rev.B23.4988-4996.Lee.1981}.
When considering Rashba SOI, electrons spins (spin up and spin down) are coupled with each other and
the total transmission $T_{pq}(E)=Tr_{r,o,s}[\Gamma_{p}
G^r_{pq}\Gamma_{q} G^a_{qp}]$, where the "${\rm Tr}_{r,o,s}$" is the trace over real space,  orbital space and spin space.

In the following, we calculate the nonlocal response, i.e., the
voltage response (detector) in top region on the current (injector)
in bottom region, which can be denoted by nonlocal
resistance $R_{23,14}$. We also calculator $R_{14,23}$, i.e., the
voltage response in bottom region on the current
in top region. $R_{23,14}$ and $R_{14,23}$
have the nearly same properties, so, we focus on $R_{23,14}$ in the
following. We apply a bias V across the injector terminals 1 and 4 to
inject current. For the detector terminals 2 and 3, the currents are
set to zero. Then use the boundary conditions $V_1=V$, $V_4=0$,
$J_2=J_3=0$, we can calculate the current $J_1 = -J_4$ and the
voltages $V_2$ and $V_3$ in the voltage probes from the
Landauer-Buttiker formula. Finally, the nonlocal resistance is given
by $R_{23,14} \equiv(V2-V3)/J_1$.

\section{numerical results and analysis}

\begin{figure}
\includegraphics[width=9cm,totalheight=6.5cm, clip=]{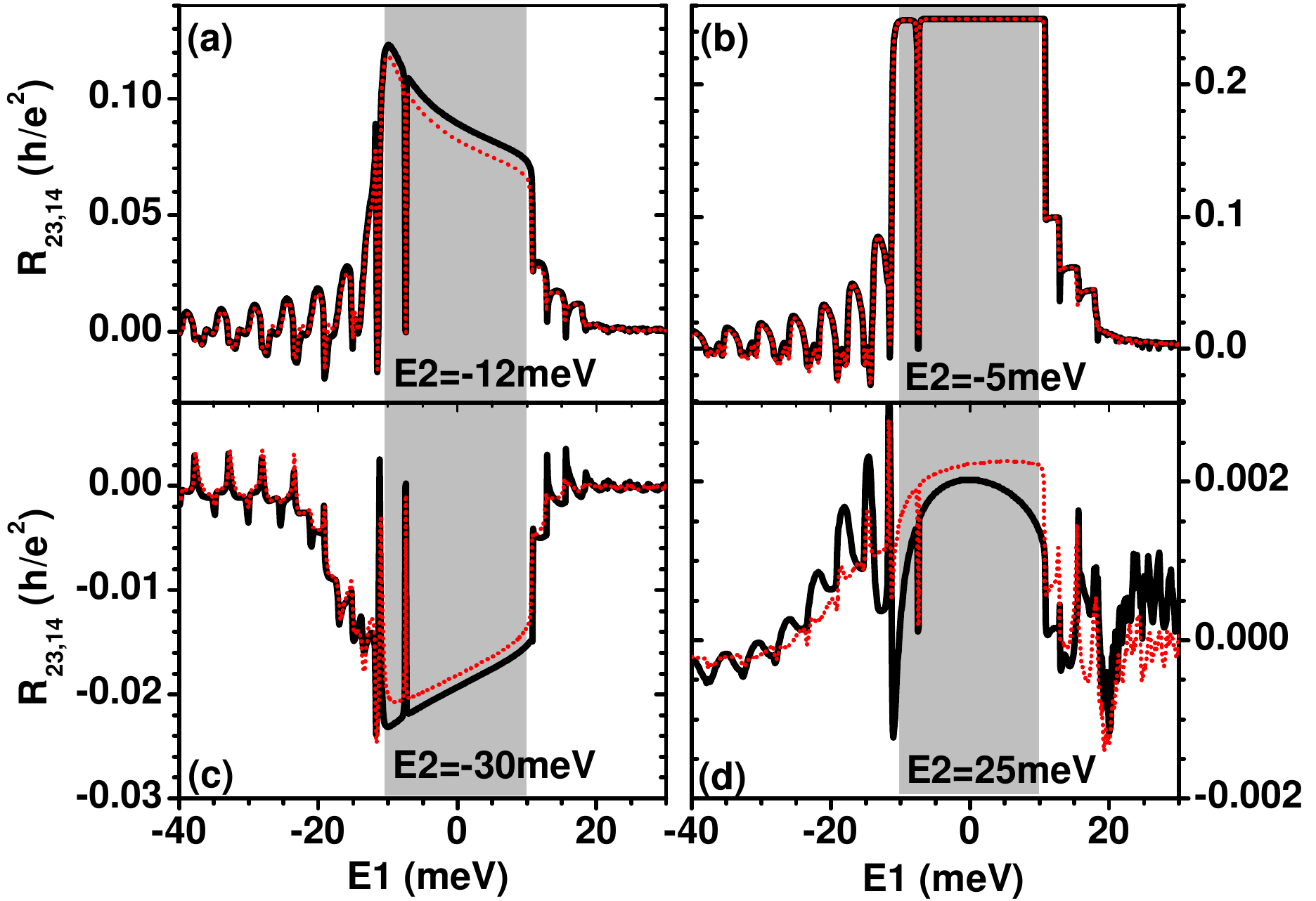} 
\caption{
$R_{23,14}$ vs on-site energy $E1$ in the injector region with Rashba SOI
$\alpha=0$ (black solid lines) and $50meV nm$ (red dotted lines).
Panels (a), (b), (c), and (d) correspond to $E2=-12meV$, $-5meV$ $-30meV$
and $25meV$, respectively, where $E2$ is the on-site energy in the detector
(top region). The gray region in panel (a,b,c,d) is the range of bulk
energy gap of injector (the sample bottom). } \label{E1}
\end{figure}

In the numerical calculation, the parameters of the HgTe/CdTe
quantum well are set as $\hbar v_F = 364.5meV nm$, $C = -686 meV
nm^2$, $D = -512 meV nm^2$, and the effective mass is taken as $m
=-10 meV$, which corresponds to the realistic quantum well with thickness
$d=7 nm$.\cite{J.Phys.Soc.Jpn.77.031007.Konig.2008} It exceeds the
critical thickness $d_c=6.3 nm$ and induces the inverted band, which
leads to the topological phase. Comparing to the inverted
InAs/GaSb/AlSb quantum well, the Rashba SOI in HgTe/CdTe quantum
well is very small and can be usually neglected in the numerical
calculation. However, in order to quantitatively estimate the effect
of Rashba SOI, we set a nonzero Rashba SOI. The strength of Rashba
SOI is set to $\alpha=50meV nm$. It is a very large value, because
the Rashba SOI in inverted InAs/GaSb/AlSb quantum well is only
$\alpha=71meV nm$,\cite{PhysicalReviewLetters100.236601.Liu.2008} while Rashba SOI in HgTe/CdTe quantum
well is much smaller than that in InAs/GaSb/AlSb quantum well. Furthermore, we set lattice constant
$a = 5 nm$, that is small enough to get a reasonable band structure.
The scattering region is shadowed in Fig.1(a). The width and length
of scattering region is set to $W=Na=750nm$ with $N=150$ and
$L=Ma=750nm$ with $M=150$. As shown in Fig.1(a), the width of the bridge
that connects the top and bottom terminals, and the distance between
top and bottom terminals, are all $250 nm$. In the numerical
calculation, we fix the Fermi energy at $E_F=0$ and change the on-site
energies $E1$ and $E2$. In fact, we can also fix $E1=E2=0$ and tuning
the Fermi energy in the two regions. They are equal in the calculation.

Based on this device, we first study the nonlocal resistance in the
clean hybrid system without the external magnetic field. With the
change of the on-site energies $E1$ and $E2$ in the injector and the detector, there will
be three different hybrid regimes: QSH-QSH, QSH-MSH, MSH-MSH. The
characters of the nonlocal response in these hybrid regimes are
depicted in Fig.2. In panel (a),(b),(c) and (d), we plot the $R_{23,14}$ vs
$E1$ in the bottom region (injector) with and without Rashba SOI for $E2=-12meV$, $-5meV$, $-30meV$ and
$25meV$, respectively.
Here, in the calculation, we set the Fermi energy $E_F=0$. And
$E2=-30meV$, $-12meV$, $-5meV$, and $25meV$ correspond to the the detector (top region) is in the n-MSH, near QSH, QSH, and p-MSH regimes, respectively. Similarly, when $E1$ changes from $-40meV$ to
$30meV$, the injector (bottom region) develops from n-MSH to p-MSH regime via QSH regime. Now,
we analyse the nonlocal properties in different hybrid structure.

\begin{figure}[t]
\includegraphics[width=9cm,totalheight=6.5cm, clip=]{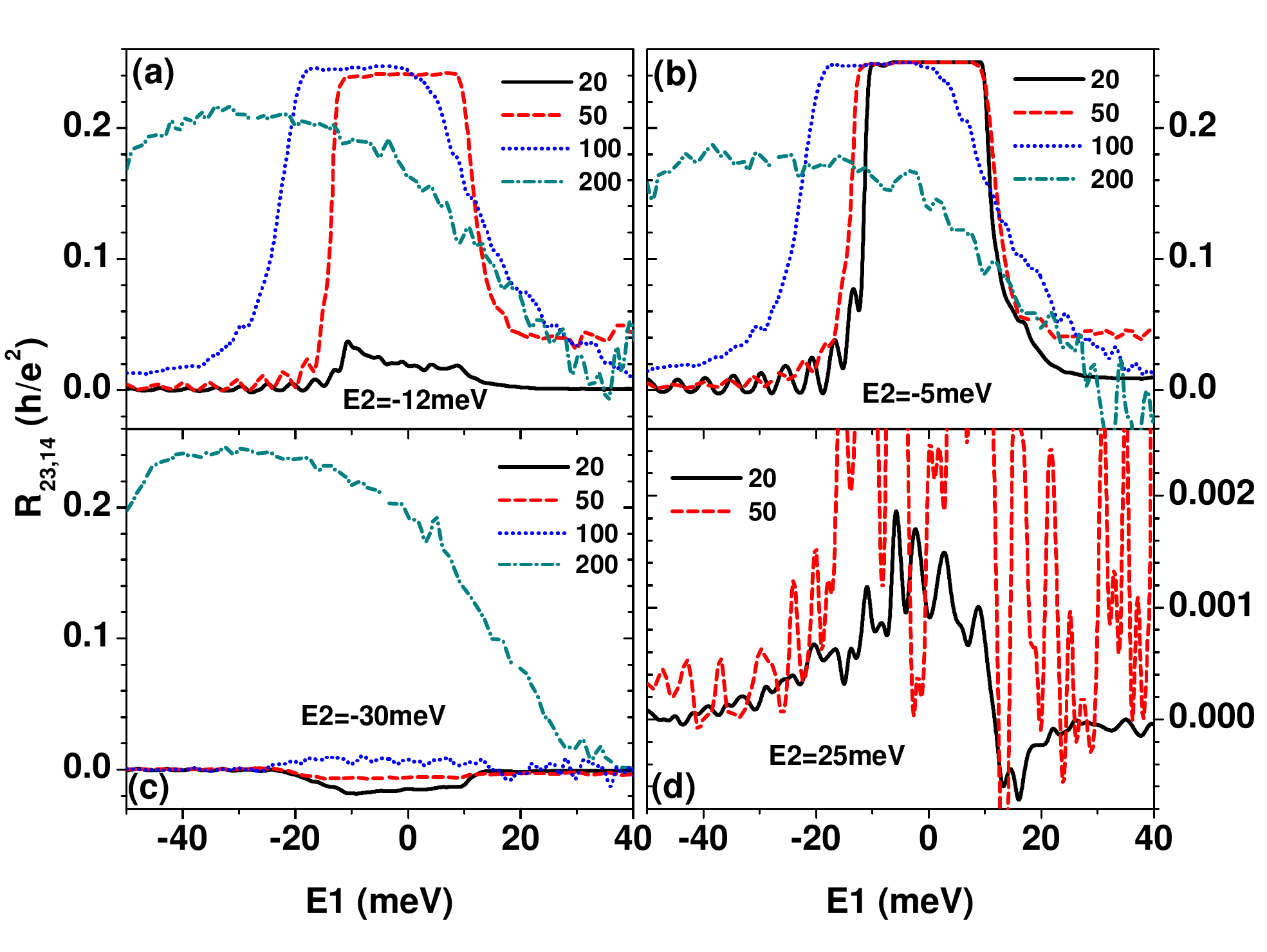} 
\caption{ (Color
online) $R_{23,14}$ vs on-site energy $E1$ in the injector region for
different disorder strength $w=20meV$, $50meV$, $100meV$, and
$200meV$. Panels (a), (b), (c), and (d) correspond to
$E2=-12meV$, $-5meV$, $-30meV$ and $25meV$, respectively. } \label{dis}
\end{figure}

From Fig.2, we can find some common characters: (1) The Rashba SOI
hardly affects the nonlocal effects of hybrid HgTe/CdTe quantum well
in variant regimes (QSH-QSH, QSH-MSH, MSH-MSH regimes, and so on).
For the system with $\alpha=0$ (black solid lines) and $\alpha=50meV
nm$ (red dotted lines), the nonlocal resistance are almost the same,
there are only slightly quantitative difference between them. So the Rashba SOI is unimportant on studying the nonlocal
effect of topological insulator. (2) No matter what the values of $E2$, $R_{23,14}$ is always biggest when the injector is in the QSH
regime, i.e., $E1\in[-10meV,10meV]$. This means the nonlocal effect
is most remarkable in the QSH regime, that is well consistent with the
experiment in Ref.[\onlinecite{NaturePhysics8.486-491.Brune.2012}].
(3) $R_{23,14}$ in the n-MSH and p-MSH regimes is small and shows
oscillating behavior, it can even be negative for some special $E1$.
Furthermore, the value of $R_{23,14}$ in the p-MSH regime is bigger than that
in the n-MSH regime, as shown in the
experiment.\cite{NaturePhysics8.486-491.Brune.2012} (4) At
$E1\approx-7.5meV$, because of the finite size effect, $R_{24,13}$ abruptly peaks or dips.
Now we analysis Fig.2 in
detail. We focus on several main regime: the QSH-QSH regime [the central
region of Fig.2(b)], the QSH-MSH regime [the left and right regions of
Fig.2(b)], and the MSH-MSH regime [the left or right panel in Figs.2(c) and 2(d)]. When the
injector and detector regions are all in the QSH regime (QSH-QSH hybrid
regime), because of the counter propagating helical edge states,
$R_{23,14}$ is biggest and is quantized at the value of
$\frac{h}{4e^2}$, regardless of the Rashba SOI [see the central region
in Fig.2(b)]. It is interpreted as follows. In the presence of the helical edge states, the
transmission coefficients are integer
$T_{12/21}=T_{23/32}=T_{34/43}=T_{41/14}=1$, we can then conclude
$V_2=\frac{2}{3}V_1+\frac{1}{3}V_4$,
$V_3=\frac{1}{3}V_1+\frac{2}{3}V_4$, and $J_1=-J_4 =(4/3)(V_1-V_4)$
from the Landauer-Buttiker formula. Then,
$R_{23,14}=\frac{V_2-V_3}{J_1}=0.25h/e^2$. Next we consider that the energy level of the detector is $E2=-12meV$. In this case, although the Fermi
energy of the detector is in the conduction band, it is very near
the band edge, $10meV$, in which the bulk states coexist with edge
states that leads to some exotic behaviors. We will call it the near QSH
regime. When the detector is in the near QSH regime and the injector
is in the QSH regime, the hybrid system will be in the near-QSH-QSH regime. In this
regime, $R_{23,14}$ is decreased by the bulk states. However, the
nonlocal effect is still strong [see Fig.2(a)] due to the helical
edge states that survive in the bulk near band edge. Beyond the
QSH-QSH regime, the MSH effects paly a role on the nonlocal transport and
$R_{23,14}$ decreases because the spin Hall effect in MSH phase is
weaker than that in QSH phase. In the QSH-nMSH and QSH-pMSH regimes
[see left and right regions of Fig.2(b)], because the eigenstates of
the MSH system are extended in the transverse
direction\cite{Phys.Rev.B73.205339.Xing.2006}, $R_{23,14}$ decreases
rapidly and oscillates around zero. The oscillating frequency is
coincident to the sub-band distribution in the conduction and valence
bands. Finally, in the MSH-MSH regime [Fig.2(c,d)], the nonlocal resistance is
induced by MSH completely, so $R_{23,14}$ is oscillating and very small.
In this case, the nonlocal resistance $R_{23,14}$ can
be negative when $E1$ is of some special values.

\begin{figure}[t]
\includegraphics[width=7.5cm,totalheight=9.5cm, clip=]{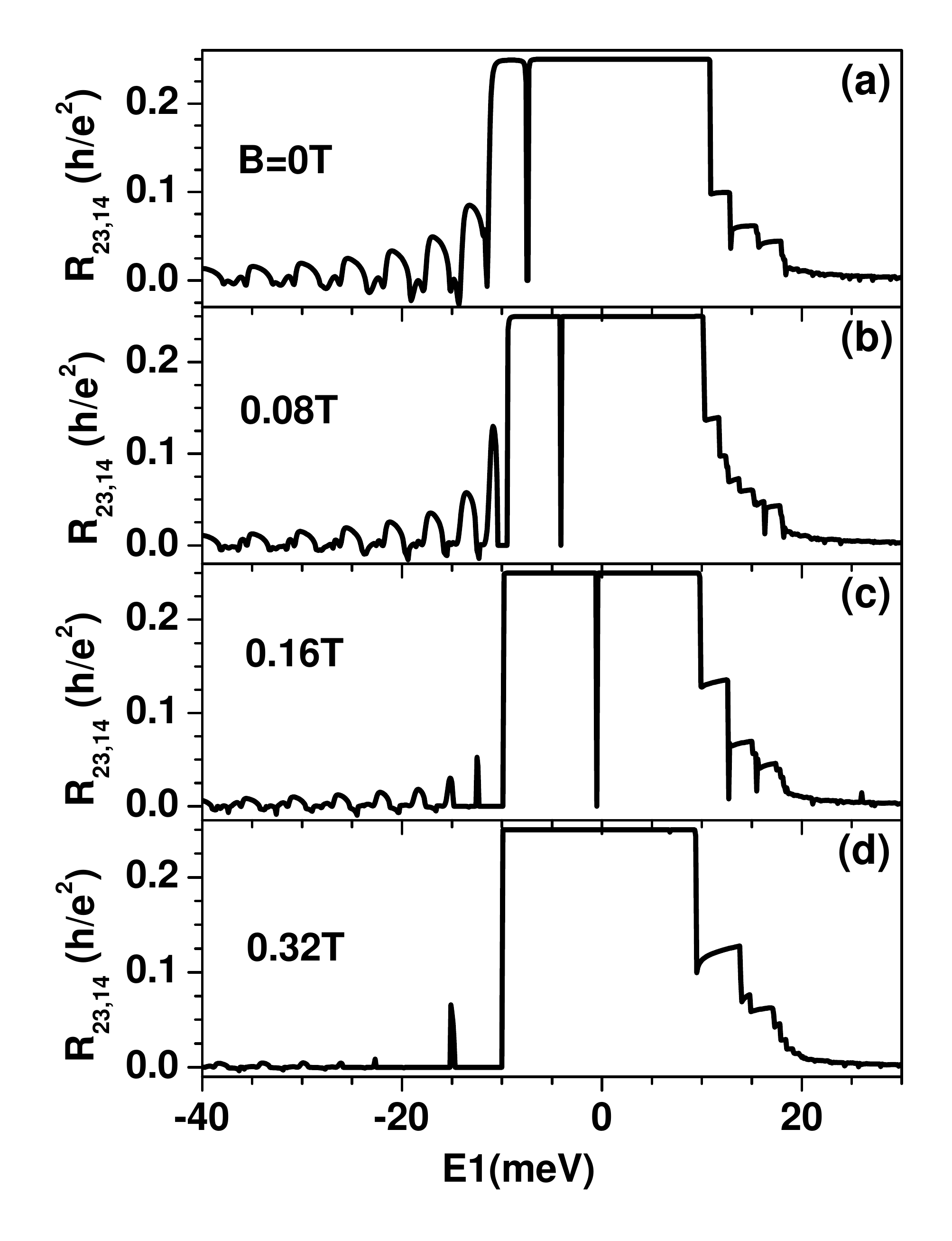}
\caption{ Nonlocal
resistance $R_{23,14}$ vs on-site energy $E1$ in the injector region.
The on-site energy in the detector region is set as $E2=-5meV$. Panels (a), (b), (c), and (d) correspond
to the magnetic field $B=0$, $0.08T$, $0.16T$ and $0.32T$, respectively.}
\label{mag1}
\end{figure}
\begin{figure*}
\includegraphics[width=12.5cm,totalheight=9.0cm, clip=]{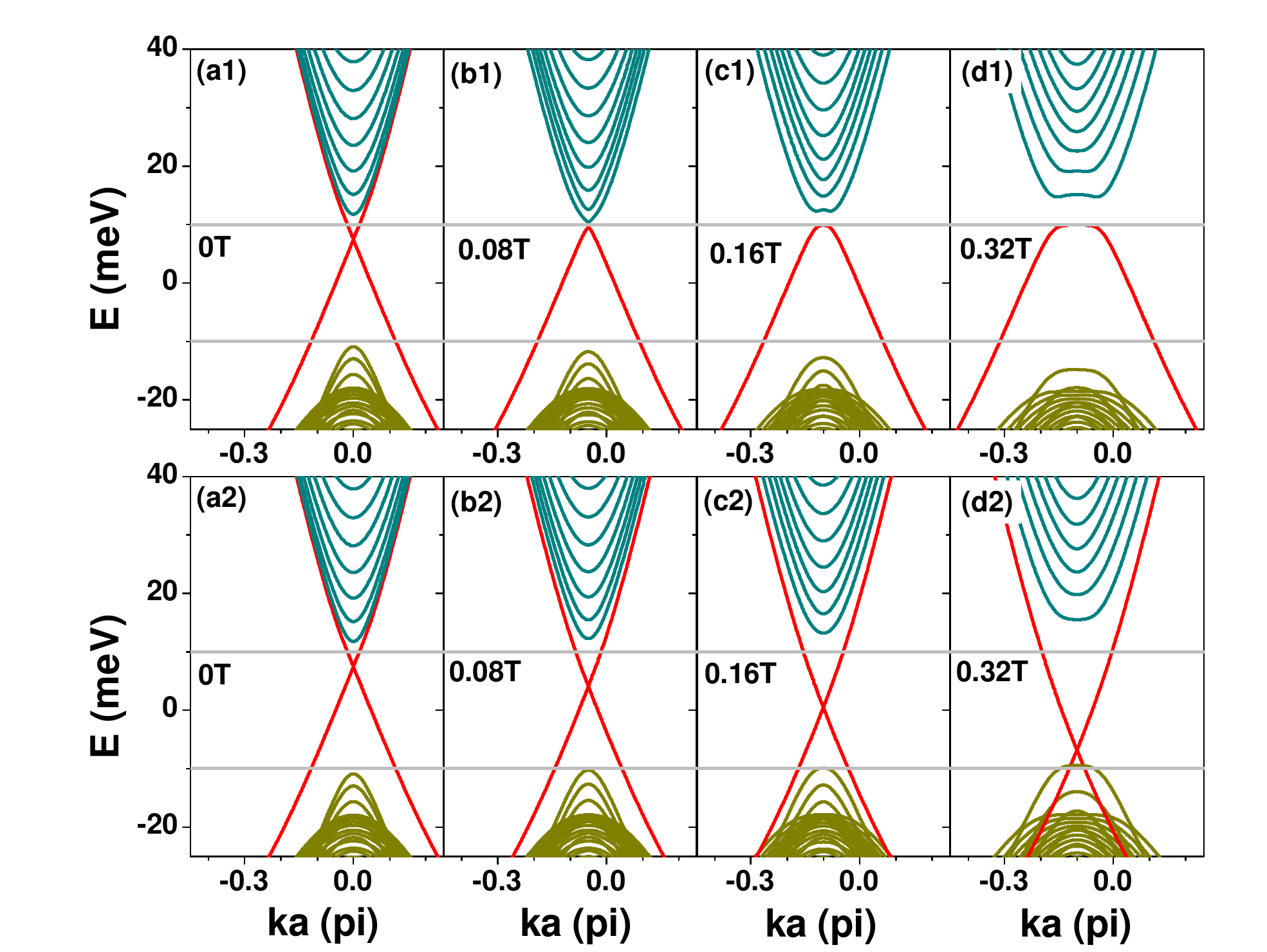}
\caption{ (Color
online) Band structure of HgTe/CdTe quantum well in the presence of
weak magnetic field. The bottom and top panels are for spin up and
spin down, respectively. Two gray lines denote the edge of band gap
in the absence of magnetic field. The on-site energy is set to zero.}
\label{band1}
\end{figure*}

Since the Rashba SOI is not important in the nonlocal effect,
only the system with $\alpha=0$ is studied in the following.
In Fig.3, we study the disorder effect of the nonlocal resistance
$R_{23,14}$ in zero magnetic field. We choose an H-shaped scattering
regions [the shadowed region in Fig.1(a)], in which the on-site
potential $w_{\mathbf i}$ is randomly distributed in the region of
$[-\frac{w}{2},\frac{w}{2}]$, with $w$ the disorder strength.
When the detector (related to $E2$) is in the QSH regime [Fig.3(b)],
it is possible for the system to be in the QSH-QSH, MSH-QSH and near-QSH-QSH regimes.
In the presence of weak disorder ($w=20meV$), $R_{23,14}$ is quantized in the QSH-QSH regime,
and the oscillating of $R_{23,14}$ in the MSH-QSH regime is strongly depressed
[see the solid lines in Fig.2(b) and Fig.3(b)].
In the moderate disorder ($w=50meV$), $R_{23,14}$ in the QSH-QSH regime is still kept but $R_{23,14}$ in QSH-MSH is increased [see red dashed line in Fig.3(b)]. Besides, due to the topological Anderson insulator
phenomenon\cite{Phys.Rev.B84.035110.Xing.2011,Phys.Rev.Lett.102.136806.Li.2009},
$R_{23,14}$ in the near-QSH-QSH regime, where the detector is near QSH regime and the injector is in the QSH regime,
is quantized by the moderate disorder [see the red dashed line in Fig.3(a)].
It means detector that is near QSH regime is now driven into the QSH regime by the moderate disorder. Then the disorder effect in QSH-QSH and near-QSH-QSH regime are almost same [see blue dotted
lines in Fig.3(a) and Fig.3(b)]. When both $E1$ and $E2$ are far away from the energy gap,
the system is in the MSH-MSH regime and $R_{23,14}$ is increased by the weak disorder [see Fig.3(d)]. When the
disorder is very strong, random scattering dominates the nonlocal
transport and $R_{23,14}$ in all the regimes are nearly the same [see the green dash dotted lines Fig.3(a,b,c)].

\begin{figure}
\includegraphics[width=8.5cm,totalheight=6.0cm, clip=]{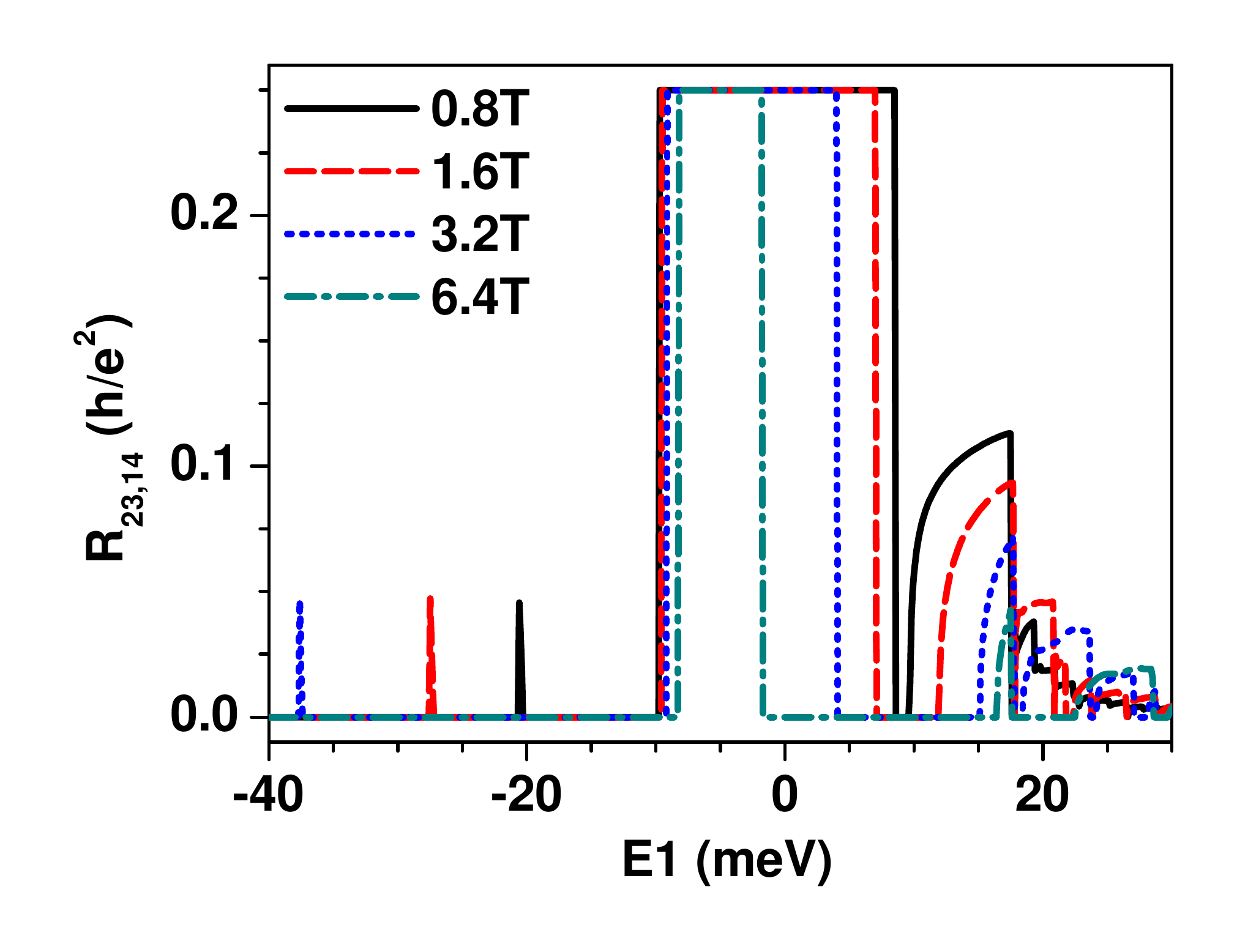}
\caption{(Color
online) Nonlocal resistance $R_{23,14}$ vs on-site energy $E1$ in the
injector region for different magnetic field. The on-site energy in detector region is set as
$E2=-5meV$.} \label{mag2}
\end{figure}
\begin{figure*}
\includegraphics[width=12.5cm,totalheight=9.0cm, clip=]{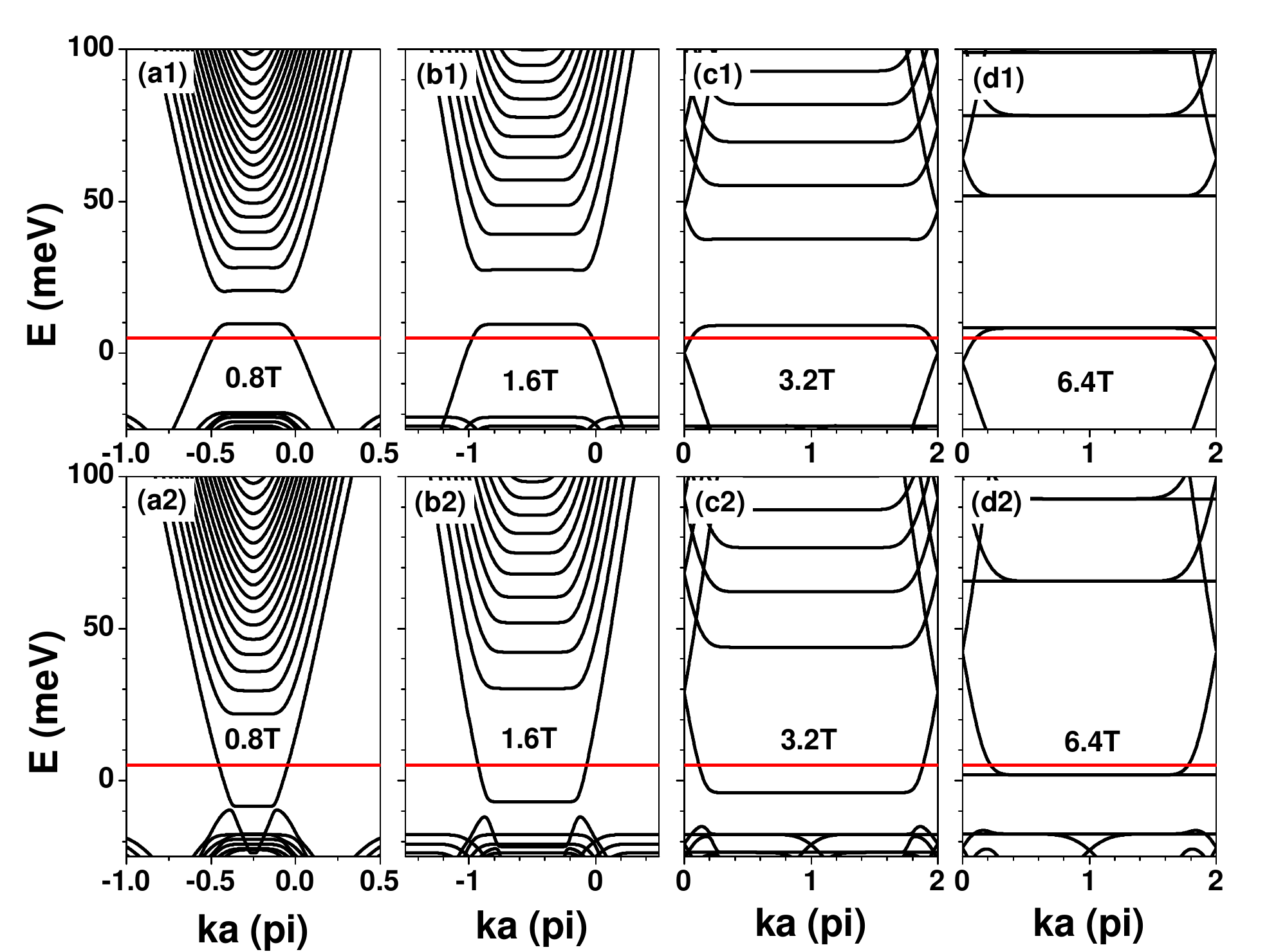}
\caption{ (Color
online) Band structure of HgTe/CdTe quantum well in the presence of
strong magnetic field. The bottom and top panels are for spin up and
spin down, respectively. The on-site energy is set to zero. The red
lines denote the position of Fermi level relative to the on-site energy.}
\label{band2}
\end{figure*}

Now, we study the nonlocal transport under the external magnetic
field. We first consider the weak magnetic field. In the HgTe/CdTe
quantum well, we are interested in the nonlocal properties of the helical
edge states. So, in the following, we consider only the hybrid system
with $E2=-5meV$, i.e., the detector is in the QSH regime. Fig.4
depicts $R_{23,14}$ vs $E1$ for different weak magnetic field $B$.
The corresponding band structures are plotted in Fig.5. From Fig.4,
we can find with increasing magnetic field $B$, the quantized
$R_{23,14}$ in the QSH-QSH regime hardly changes and $R_{23,14}$ in the QSH-nMSH regime is suppressed gradually.
This can be explained as follows. As we know, the quantized $R_{23,14}$ in the QSH-QSH regime arises from the helical edge states, while $R_{23,14}$ in the QSH-nMSH regime is dominated by the extended eigenstates\cite{Phys.Rev.B73.205339.Xing.2006} in the MSH system. Considering the transport process, the helical edge states propagate only along the {\sl geometric edge}, while the extended MSH states can propagate in the {\sl whole scattering region}. Thus, when the external magnetic field is added (in all regions including the four leads and the scattering region), it can drastically affect $R_{23,14}$ in the QSH-MSH regime but hardly changes the quantized $R_{23,14}$. All in all, in the presence of weak magnetic field, it is the helical edge states and the extended MSH states that induce the unchanged quantized $R_{23,14}$ in the QSH-QSH regime and drastically suppress $R_{23,14}$ in the QSH-nMSH regime, as in the weakly disordered system.
Besides, in Figs.4(b), 4(c), and 4(d), we can see that in some energy regions, e.g.,
$E1$ near $-10meV$ as in Fig.4(b), $E1\in[-15meV,-10meV]$ as in
Fig.4(c) and $E1\in[-25meV,-10meV]$ as in Fig.4(d), except some abrupt peaks, $R_{23,14}$ is
zero. This can be interpreted as follows.
When the magnetic field
increases, the flat Landau level forms gradually in the conduction
band. The topological edge state near
the conduction band gap with spin down is first broken by the magnetic
field, because its chirality, i.e., the rotation direction along the
scatter edge, which can be characterized by Chern number
C\cite{Phys.Rev.Lett.71.3697--3700.Hatsugai.1993,Phys.Rev.Lett.97.036808.Sheng.2006},
is opposite to the edge state induced by magnetic field. The
opposite chirality also induces the slight dip in Landau levels [see the
top panels in Fig.5], which leads to the abrupt peaks in the conduction
band\cite{Phys.Rev.B85.125401.Chen.2012} as shown in Fig.4(c) and (d).
On the other hand, for the carriers with spin up, the helical edge state near
topological band gap is kept [see bottom panels in Fig.5] because
its chirality is same to the Landau edge state.
Combining the Landau gap in the system with spin down and the topological edge state in the system with spin up, the system is equal to a quantum Hall system, in which the edge state is unidirectional, since the edge state
contributed by spin down is broken by the magnetic field. As a result,
the chemical potential $V_2$ and $V_3$ of the detector terminals are
determined by one of the adjacent source terminals. It means the chemical
potential $V_2=V_1$ and $V_3=V_2$, which leads to zero nonlocal
resistance $R_{23,14}$. Finally, there are also abrupt dips in
quantized $R_{23,14}$, as interpreted in the zero magnetic field
case, which comes from the finite size effect.

Next, we consider the nonlocal effect in the strong magnetic field.
In this case, the edge states for spin up and spin down all develop
into flat Landau levels. Fig.6 plots the nonlocal resistance
$R_{23,14}$ vs on-site energy $E1$ in strong magnetic field $B$.
Three characters are found: (1) Although $R_{23,14}$ is quantized in the bulk gap of [-10meV,10meV], the
quantized range gradually shrinks with increasing of $B$. (2)
Except several abrupt peaks, $R_{23,14}$ is nearly zero in the whole
conduction band. (3) $R_{23,14}$ in the valence band is zero in some
region, this region expands with increasing of $B$. To
understand these behaviors, we need to analysis the band influence
induced by strong magnetic field.

As we know in the HgTe/CdTe topological insulator, the chirality of
the edge state induced by the inverted band is identical for electron and
hole but opposite for carriers with spin up and spin down. On the
other hand, in the strong magnetic field, the chirality of Landau
edge state induced by the magnetic field is same for carriers with spin
up and spin down but opposite for electron and hole. For all four
classes of carrier, i.e., electrons and holes with spin up and spin
down (signed as $e\uparrow$, $e\downarrow$, $h\uparrow$,
$h\downarrow$), when the two type edge states,i.e., edge state induced by
inverted band and magnetic field (signed by `E' and `B') have same
chirality, they can coexist and boost. Otherwise, they are annihilated with
each other. In our system, concerning the chirality of two types of
edge states,
$C_{h\uparrow,E}=C_{e\uparrow,E}=C_{e\uparrow,B}=C_{e\downarrow,B}=1$,
$C_{h\uparrow,B}=C_{h\downarrow,B}=C_{h\downarrow,E}=C_{e\downarrow,E}=-1$,
the edge states of $e\uparrow$ and $h\downarrow$ are strengthened, and those of
$h\uparrow$ and $e\downarrow$ are destroyed in the strong magnetic
field, as shown in Fig.7. However, since the edge states induced by the
inverted band is more robust in the energy close to the conduction
band, the destroy is more gentle (see top panels in Fig.7). So, in
the strong magnetic field, the topological edge states induced by
inverted band are gradually destroyed in the region of
$E\in[-10meV,0]$ and nearly kept in the region of $E\in[0,10meV]$ as
shown in Fig.7. And in the region of $E1\in[0,10meV]$ in Fig.6 (corresponding to $E\in[-10meV,0]$ in Fig.7),
one of two helical edge states gradually disappears and the system can be regarded as a quantum Hall system with
$R_{23,14}$ being zero. Then we demonstrate the first character in the last paragraph.
Furthermore, in the presence of strong magnetic field, the Landau levels are completely formed in the conduction band.
In the deep conduction band, the topological edge state induced by
inverted band does not work and the unidirectional Landau edge states dominate the transport procession,
leading to the zero $R_{23,14}$ in the conduction band of $E1<-10meV$. Besides zero
$R_{23,14}$, there are also some abrupt peaks in the conduction band,
it is because of the slightly dip in Landau level.
This is for the second
character. Finally, in the valence band, although the edge states of
spin up are destroyed by the magnetic field, Landau levels have not
been completely formed and there still exist extend states between
Landau levels, which leads to the nonzero $R_{23,14}$ in valence
band. Except these extent states, Landau levels are entirely gapped and
$R_{23,14}$ is then zero. This interprets the third character in
Fig.6.

Besides the three characters depicted in Fig.6, we can also expect
when $B$ is large enough, the edge states of $h\uparrow$ and $e\downarrow$ are also completely destroyed, and the topological
gap $[-10meV,10meV]$ in these system will become the real gap in
which both the bulk and edge state are all absent.\cite{Phys.Rev.B85.125401.Chen.2012} Then, the
quantized region of $R_{23,14}$ will disappear entirely.
Furthermore, when $B$ is very large, in the conduction and valence
bands, the Landau edge state will dominate the transport and the system
come into the quantum Hall regime, in which $R_{23,14}$ is strictly
zero in any hybrid regime. So, the nonlocal resistance in the H-shaped
hybrid HgTe/CdTe quantum well will completely disappear in very large magnetic
field.

\begin{figure}
\includegraphics[width=7.0cm,totalheight=5.0cm, clip=]{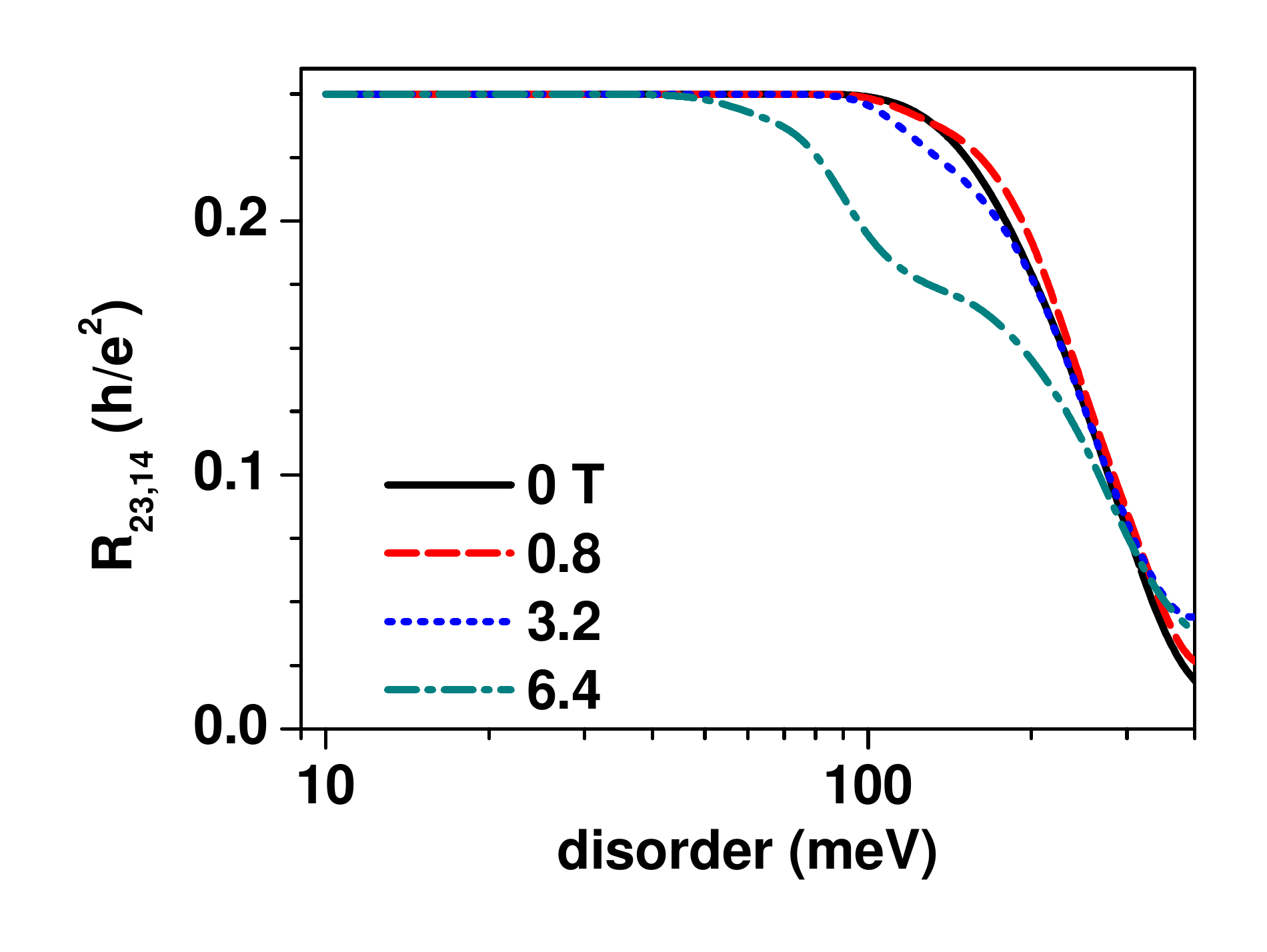}
\caption{ (Color
online) Nonlocal resistance $R_{23,14}$ vs disorder strength $w$ for
different magnetic field. The on-site energies of the injector and detector regions are
set as $E1=E2=-5meV$.} \label{magdis}
\end{figure}

In the following, we study the disorder effect on the nonlocal
resistance in the QSH-QSH regime in the presence of strong magnetic
field. We set the on-site energies in the injector and the detector as
$E1=E2=-5meV$ and Fermi energy $E_F=0$, which equal to $E_F=5$ and
$E1=E2=0$. With these parameters, helical QSH edge state can be
maintained at very large magnetic field $B=6.4T$ [see Fig.7(d1) and (d2) in which $E_F=5$
and $E=0$]. However, because the magnetic field destroys the edge
state in the band edge, it weakens the robustness of the helical edge
state, especially for the edge states located near the gap edge, as
shown in Fig.8, in which $R_{23,14}$ vs disorder strength for
different magnetic field is plotted. It can be seen when $B=6.4T$,
the quantized $R_{23,14}$ is decreased at small disorder strength of
$w\approx40meV$, while for $B=0$, $0.8T$ and $3.2T$, quantized
$R_{23,14}$ can be maintained even at $w=100meV$. This is because at
$B=6.4T$ [Fig.7(d1) and (d2)], the helical band gap is violently shrunk into the region of $[2meV,8meV]$ and
$E_F=5meV$ is very close to the gap edge. For the magnetic field $B\leq3.2T$, the band gap is wider of about $[-5meV,10meV]$. In this case, the Fermi energy is far away from the gap edge and the
nonlocal resistance can be maintained in the strong disorder scattering.

\section{conclusion}

In summary, we have investigated the nonlocal transport in a
H-shaped hybrid HgTe/CdTe quantum well. Three hybrid regime are
considered, QSH-QSH regime, QSH-MSH regime and MSH-MSH regime. It is
found in the QSH-QSH regime, the spin polarized edge states
denominate the transport procession, the nonlocal effect is most
strong, and the nonlocal resistance $R_{23,14}$ is quantized in the
value of $0.25\frac{h}{e^2}$. While in the QSH-MSH device, due to
the extended states in MSH effect, the nonlocal resistance is
oscillating and much smaller than in the QSH-QSH device. In the
MSH-MSH regime, the nonlocal effect nearly disappear, $R_{23,14}$
can only reach the order of $0.01\frac{h}{e^2}$.
In the presence of the disorder, the quantized nonlocal
resistance in the QSH-QSH device is robust
because of the time reversal protected edge states. Near the QSH-QSH
regime, the nonlocal resistance is enhanced and quantized due to the
topological Anderson insulator phenomenon. While in the QSH-MSH and MSH-MSH
device, the oscillated nonlocal resistance is strongly restricted by
disorder. Finally, the magnetic field effect is investigated. It is
found the quantized nonlocal resistance in QSH-QSH regime can't be
affect by weak magnetic field, but the nonlocal resistance of
QSH-MSH device is smoothed out by $B$. With the increasing of $B$, the Landau level forms,
counter propagating edge states are replaced by chiral edge state, so
the region of quantized nonlocal resistance decrease and disappear
finally. All these aforementioned nonlocal features can be derived
from the special band structure of HgTe/CdTe quantum well. It is
the unique property of HgTe/CdTe quantum well in topological
insulator phase, it can be regarded as the fingerprint of the
helical QSH edge states in 2D topological insulator.

$${\bf ACKNOWLEDGMENTS}$$
We gratefully acknowledge the financial support from the National
Natural Science Foundation of China (No. 11174032 and 11274364),
NBRP of China (2012CB921303), and Ph.D. Programs Foundation of
Ministry of Education of China (No.20111101120024). We thank Ai-Min Guo for help discussions.
\\
$*$ xingyanxia@bit.edu.cn \\
$\dagger$ qfsun@pku.edu.cn

\end{document}